\documentclass[prd, twocolumn, eqsecnum,nofootinbib,floatfix]{revtex4}
\usepackage{amsmath,amssymb,amsfonts,epsfig,graphicx,euscript}
\usepackage{graphics}
\usepackage{float}
\newcommand{\bse}{\begin{subequations}}
\newcommand{\ese}{\end{subequations}}
\newcommand{\be}{\begin{equation}}
\newcommand{\ee}{\end{equation}}
\newcommand{\bea}{\begin{eqnarray}}
\newcommand{\eea}{\end{eqnarray}}
\newcommand{\ba}{\begin{array}}
\newcommand{\ea}{\end{array}}

\begin{document}

\begin{flushright}
\end{flushright}
\begin{flushright}
\end{flushright}
\hfill%

\title{Gravity Induced Quantum Interference on Gravitational Wave Background}
\author{ Mohammad A. Ganjali\footnote{ganjali@theory.ipm.ac.ir}}
\affiliation{Department of Physics, Kharazmi University,\\P. O. Box
31979-37551, Tehran, Iran}
\author{Zainab Sedaghatmanesh\footnote{std\_sedaghatmanesh@khu.ac.ir}}
\affiliation{Department of Physics, Kharazmi University,\\P. O. Box
31979-37551, Tehran, Iran}
\begin{abstract}
Gravity-induced quantum interference is an experiment that exhibits how a gravitational effect appears in quantum mechanics \cite{induced}. In this famous experiment, gravity was added to the system just classically. In our study, we do the related calculations on a gravitational wave background. We realize that the effect of gravitational wave would be detectable in this quantum mechanical effect.\\
\end{abstract}
\maketitle
\section{introduction}
\begin{figure}[H]\center
\label{path}
\includegraphics[scale=0.45]{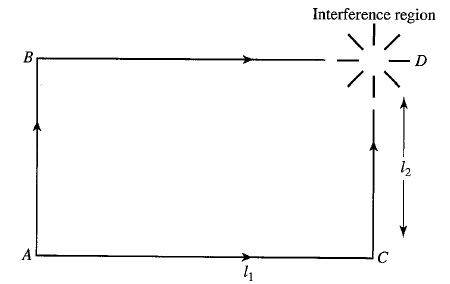}
\caption{a general view of the experiment}
\end{figure}

We learned from quantum mechanics that a change in potential
\bea
V(\textbf{x})\rightarrow  V(\textbf{x})+V_0(t)
\eea
would result a change in phase of state as
\bea
|\alpha,t_0;t>\rightarrow \exp\left[-i\int_{t_0}^{t}{dt'\frac{V_0(t')}{\hbar}}\right]|\alpha,t_0,t>.
\eea
Due to the above fact, in a famous and very interesting experiment in 1975, known as \textit{gravity-induced quantum interference}\cite{induced}\cite{Abele}\cite{Abele:2012dn}\cite{sakuraei}, a nearly mono-energetic beam of thermal neutrons was split into two parts and then brought together by silicon\footnote{Due to Sakuraei \textit{"Beautiful art of neutron interferometry!"}} as in Figure (\ref{path}) in order to exhibit gravitational effect in quantum gravity. Since the size of wave packet is much smaller than the macroscopic dimension of the two loops, it was considered classical paths for neutrons. Consider $A\rightarrow B\rightarrow D$ and $A\rightarrow C\rightarrow D$ paths lie in a rectangular plane which is rotated around segment $AC$ by angle $\delta$ and this configuration is placed on a gravitational background with potential
\bea
V(y)=-gy
\eea
where $g=9.8m/s^2$. Here, it was assumed that the $AC$ is a reference line for potential in which $V(AC)=0$.
Since neutrons examine different potentials for the lines $AC$ and $BD$, one expects a phase difference for wave packets of these two paths as
\bea
\label{classic}
exp\left[-i\Delta\phi_0\right]&=&exp \left[ \dfrac{-im_ng l_1l_2}{\hbar v_{0x}}\sin\delta\right]\nonumber\\
&=&exp \left[ \dfrac{-im^2_ngl_1l_2\bar{\lambda}}{\hbar^2}\sin\delta\right]
\eea
and that an observable interference would occur. Here $l_1$ and $l_2$ are the lengths of $AC=BD$ and $AB=CD$ arms respectively. $\bar{\lambda}$ is the neutron's De Broglie wavelength and $m_n$ is the neutron mass. For $\bar{\lambda}=1.42{\AA}$ and $l_1l_2=10 cm^2$, one obtains $\frac{m^2_ngl_1l_2\bar{\lambda}}{\hbar^2}=55.6$ and that rotating the loop plane would imply a series of maxima and minima in the interference intensity. Surprisingly, such a phase shift due to gravity has been verified within $1\%$.

Analogous to this quantum experiment with classical gravity, now we aim to calculate the phase difference of neutrons when the experiment is done on a gravitational wave (GW) background. In fact, due to GW, the paths of neutrons are slightly changed and the travel time would be different from that of non-GW case. Therefore, one expects a correction to the phase difference. Although, gravity induced phase difference in the framework of general relativity has been studied in several works \cite{Ryder1},\cite{Galiautdinov:2017lcx} but, in this work we consider the especial case of gravitational wave metric.

Our computations in this paper are presented in the next section \textbf{II} where we calculate wave function phase difference on the background of a gravitational wave. In particular, we will do our analysis for two different modes of gravity wave i.e $f_+$ and $f_{\times}$ in subsections \textbf{A} and \textbf{B} of section \textbf{II}, separately. In summary section, we discuss about our main results on the possibility of detecting a GW-induced phase difference effect in neutron interferometer experiment.
\section{the phase difference induced by gravitational waves}
The gravitational wave we consider here propagates in the $z$ direction and the line element in this background is given by \cite{Chen:2016isk}\cite{Cai:2017cbj}
\bea
\label{metric}
ds^2=c^2dt^2&-&dz^2
-\left(1-f_{+}(z-ct)\right)dx^2\\
&-&\left(1+f_{+}(z-ct)\right)dy^2
+2f_{\times}(z-ct)dxdy,\nonumber
\eea
where $ f_+ $ and $ f_\times $ are plus and cross polarization of the GW.
The problem is the same as before but now one should note that in presence of gravitational waves the lengths of paths from which the neutrons pass (and consequently the time) are altered \cite{maggiore}. In order to calculate the paths and time taken to reach the detector we need to solve geodesic equations for neutron particles on the background (\ref{metric}). Considering the geodesic equation
\begin{align}
\ddot{x}^{a}+\Gamma^{a}_{bc}\dot{x}^{b}\dot{x}^{c}=-g^{ab}\partial_{b} V,
\end{align}
we will do this in two subsections one with the condition $ f_{\times}=0 $ and the other with $ f_+=0 $.
\subsection{$f_+$ induced phase difference}
Taking $ f_+=a \cos(\omega t-kz) $ and $ f_{\times}=0 $ for simplicity with $ V=-gy $ the geodesic equation for $x$ and $y$ components for $BD$ arm become
\bea
\ddot{x}+2\Gamma^{x}_{xt}\dot{x}&=&0,\\
\ddot{y}+2\Gamma^{y}_{yt}\dot{y}&=&-g g^{yy}
\eea
respectively. Integrating these equations and keeping the terms up to linear order of $"a"$ give us
\begin{align}
\label{x}
x-x_0&=v_{0x}[(1-a)t+\dfrac{a}{\omega}sin(\omega t)]+{\cal O}(a^2),
\end{align}
\begin{align}
\label{y}
y-y_0&=\dfrac{1}{2}gt^2-\dfrac{ga}{\omega^2}[\cos(\omega t)+\omega t\sin(\omega t)-1],
\end{align}
where $ v_{0x} $ is the initial velocity at $"BD"$ arm and we have taken $ v_{0y}=0 $. Here $(x_0,y_0)=(0,l_2)$ and we consider $x-x_0=l_1$ but $y$ would be determined by classical kinematics where now the GW effect is added.\\
Then, it remains to find the travel time from $B$ to $D$. To do so we should solve the above equations to reach $T$. For this aim one may combine these two coupled equations by considering that at time $T$ we have $x-x_0=l_1$ and $y-y_0=0$ which leads to a quadratic equation for $T$. But, instead of such method, we expect that the traveling time should have the following form
\bea\label{T}
 T=T_0+\alpha.a +{\cal O}(a^2),
\eea
where $ T_0=\dfrac{l_1}{v_{0x}} $. Then, inserting this in equation (\ref{x}) we easily get
\bea
\alpha = T_0\left(1-sinc(\omega T_0)\right).
\eea
where $sinc\;x=\frac{\sin{x}}{x}$.

Finally, the wave packet arriving at $D$ via the path $"BD"$ has a phase change
\begin{widetext}
\bea\label{phase+}
exp\left[-\dfrac{im_n}{\hbar} g\sin \delta.y.T\right]=
exp\left\lbrace -\dfrac{im_n}{\hbar} g\sin\delta \left[  l_2+\dfrac{1}{2}gT^2-\dfrac{ga}{\omega^2}\left( \cos(\omega T)+\omega T \sin(\omega T)-1\right) \right] . T
\right\rbrace
\eea
\end{widetext}
and that this phase change from $"ABD"$ path is the phase difference relative to that when the wave packet arriving at $D$ via path $"ACD"$.

As it is seen the first term in (\ref{phase+}) is the same as (\ref{classic}) and there are some correction terms which come from gravitational waves (and also classical kinematics). Recalling that $T$ is given by (\ref{T}) and expanding the various terms in (\ref{phase+}) we get
\begin{widetext}
\bea
\label{fi}
-i\Delta\phi &=&-i(\phi_{ABD}-\phi_{ACD})=-i(\Delta\phi_0+\Delta\phi_{corr})=
-i\Delta\phi_0\left(1+\delta\right)\\
&=&-i\Delta\phi_0 \left\lbrace   1+\frac{1}{2}\frac{g}{l_2}T_0^2
+ a\left[1-sinc(\omega T_0) +\frac{1}{2}\frac{g}{l_2}T_0^2\left(sinc^2(\frac{\omega T_0}{2})-5 sinc(\omega T_0)+3\right)\right] +{\cal O}(a^2)\right\rbrace,\nonumber
\eea
\end{widetext}
where we have defined
\bea
\delta &=&\delta_{class}+\delta_{GW}\\
\delta_{class}&=&\frac{1}{2}\frac{g}{l_2}T_0^2\\
\delta_{GW}&=&a\left[f(\omega T_0)
+\delta_{class}g(\omega T_0)+{\cal O}(a^2)\right]\nonumber
\eea
where
\bea
f(\omega T_0)&=&1-sinc(\omega T_0)\\
g(\omega T_0)&=&sinc^2(\frac{\omega T_0}{2})-5 sinc(\omega T_0)+3\nonumber
\eea
which $\delta_{class}$ is a correction term that just comes from the classical trajectory of a point particle
and $\delta_{GW}$ is the correction coming from GW effect.

Now, in order to discuss the above result some preliminary comments are in order:

First, we should say that in this study, as we discussed earlier, we would like to compute the correction to phase difference of that of the original experiment \cite{induced} where in that experiment we had $\frac{m^2_ngl_1l_2\bar{\lambda}}{\hbar^2}=55.6$. Equivalently,
\bea\label{ll}
\frac{l_1l_2}{v_{0x}}=l_2T_0=55.6\frac{\hbar}{m_ng}=10^{-7}
\eea
That is, in our study, we will consider $l_1, l_2$ and $v_{0x}$ as free parameters but any changes in these parameters should satisfy (\ref{ll}). In other words, $\Delta\phi_0$ should not change. Moreover, we note that the parameters $a$ and $\omega$ are also free but, these are not in our control at least up to now. Beside, due to perturbative general relativity we have $a\ll 1$.

Second, in experiment \cite{induced}, the term $\delta_{class}\simeq 10^{-7}$ was considered negligible. Therefore, for detecting a distinguished gravitational wave effect, i.e. the $\delta_{GW}$ term of (\ref{fi}), we should consider phenomenon that the GW effect is greater than the classical effect.

Having this considerations in mind,
let us discuss about (\ref{fi}) at the following regimes:
\begin{itemize}
\item $\omega T_0\gg 1$: In this case we have $f\mapsto 0$ and $g\mapsto 3$ and that
\bea
\delta=\delta_{class}(1+3a)+a
\eea
Again, since $a\ll 1$, GW effect does not have a chance to be visible.
\item $\omega T_0\ll 1$: In this limit, we have $f\mapsto 0$ and $g\mapsto -1$ and that
\bea
\delta=\delta_{class}(1-a).
\eea
Then, since $a\ll 1$, thus GW effect would not be visible.
\end{itemize}
At the end, when $f_+$ is turned on, it seems that detecting the gravitational wave effect on phase difference would be very hard.

Although, it would be notable that since the functions $f$ and $g$ are periodic functions with period $2\pi$, one may imagine an experimental setup where such oscillating property shows the presence of gravitational wave effect on quantum interference of two neutron beams!
\subsection{$  f_{\times}$ induced phase difference}
Now, in order to complete our discussion we take $ f_{\times}=b \cos(wt-kz) $ and $ f_+=0 $. Writing the geodesic equation for $x$ and $y$ components gives
\begin{align}
\ddot{x}+2\Gamma^x_{xt}\dot{x}+2\Gamma^x_{yt}\dot{y}=-g g^{xy},
\end{align}
\begin{align}
\ddot{y}+2\Gamma^y_{yt}\dot{y}+2\Gamma^y_{xt}\dot{x}=-g g^{yy}.
\end{align}
As it is seen $x$ and $y$ equations have been coupled due to $ f_{\times} $ related terms. Taking terms linear order in $b$ and  defining $ z=x+y $ and $ \overline{z}=x-y $ to combine the two equations finally leads to
\bea
\label{x1}
x-x_0&=&v_{0x}t+\dfrac{bg}{\omega^2}\left[ \cos(\omega t)+\omega t \sin(\omega t)-1\right]+{\cal O}(b^2),\nonumber\\\\
y-y_0&=&\dfrac{1}{2}gt^2+\dfrac{v_{0x}b}{\omega}\sin(\omega t),
\eea
in which we have taken $ v_{0y}=0 $. As before, we take
\begin{align}
T=T_0+\beta .b+{\cal O}(b^2),
\end{align}
where $ T_0=\dfrac{l_1}{v_{0x}} $. Inserting this in equation (\ref{x1}) gives
\begin{align}
\beta &=\frac{1}{2}\dfrac{g}{l_{1}}T_0^3\left(sinc^2(\frac{\omega T_0}{2})-2sinc(\omega T_0)\right) \nonumber\\ .
\end{align}
Consequently the phase change becomes
\begin{widetext}
\bea\label{phase++}
exp\left[-\dfrac{im_n}{\hbar} g\sin \delta.y.T\right]=
exp\left\lbrace -\dfrac{im_n}{\hbar} g\sin \delta .\left[l_2+\dfrac{1}{2}gT^2+\dfrac{v_{0x}b}{\omega}\sin(\omega T)\right] .T\right\rbrace
\eea
\end{widetext}
Inserting the expression for $T$, we finally get
\begin{widetext}
\bea
\label{fi2}
-i\Delta\phi &=&-i(\phi_{ABD}-\phi_{ACD})=-i(\Delta\phi_0+\Delta\phi_{corr})=
-i\Delta\phi_0\left(1+\delta\right)\\
&&=-i\Delta\phi_0 \left\lbrace 1+\dfrac{1}{2}\dfrac{g}{l_2}T_0^2+b\left[\dfrac{l_1}{l_2}sinc(\omega T_0)+\frac{1}{2}\dfrac{g}{l_1}T_0^2\left(1+\dfrac{3}{2}\dfrac{g}{l_2}T_0^2\right)\left(sinc^2(\frac{\omega T_0}{2})-2sinc(\omega T_0)\right)\right]\right\rbrace\nonumber
\eea
\end{widetext}
where we have again defined
\bea
\delta &=&\delta_{class}+\delta_{GW}\\
\delta_{class}&=&\frac{1}{2}\frac{g}{l_2}T_0^2\\
\delta_{GW}&=&a[\frac{l_1}{l_2}f(\omega T_0)
+\frac{l_2}{l_1}\delta_{class}\left(1+3\delta_{class}\right)g(\omega T_0)+{\cal O}(a^2)]\nonumber
\eea
where
\bea
f(\omega T_0)&=&sinc(\omega T_0)\\
g(\omega T_0)&=&sinc^2(\frac{\omega T_0}{2})-2sinc(\omega T_0)\nonumber
\eea
Here, again, we discuss about (\ref{fi2}) at the following regimes:
\begin{itemize}
\item $\omega T_0\gg 1$: In this case we have $f\mapsto 0$ and $g\mapsto 0$ and that
\bea
\delta=\delta_{class}
\eea
In this case, we will not detect gravitational effect anyway.
\item $\omega T_0\ll 1$: In this limit, we have $f\mapsto 1$ and $g\mapsto -1$ and that
\bea
\delta=\delta_{class}+a\left(\frac{l_1}{l_2}+\frac{l_2}{l_1}\delta_{class}(1+3\delta_{class})\right).
\eea
Here, there exist a chance in which a visible gravitational wave signal would be detected. In fact, one may consider two regimes for this aim. First take $l_1\gg 1, l_2\ll 1$ and $v_{0x}\gg 1$ so that $T_0$ and $l_1l_2=unchanged$. Then, $\delta=\delta_{class}+\frac{l_1}{l_2}a$. Similarly, taking $l_1\ll 1, l_2\gg 1$ and $v_{0x}\ll 1$ so that again $T_0$ and $l_1l_2=unchanged$ then, $\delta=\delta_{class}(1+\frac{l_2}{l_1}(1+3\delta_{class})a)$. In both cases we would have a chance to detect a visible gravitational wave effect.
\end{itemize}.

\section{Summary and conclusion }
In this work, we considered the famous gravity-induced quantum interference experiment on a gravitational wave background. We solved the geodesic equations for the paths from which neutrons pass to reach the detector (figure\ref{path}) and then by calculating the GW-altered time for reaching the detector, we finally computed the corrected phase difference of original experiment in two cases of plus and cross GW-polarizations.

The final results of our calculations are summarized in \ref{fi} and \ref{fi2} formulas, which are made of two main parts $\delta_{class}$ and $\delta_{GW}$. $\delta_{GW}$ is the correction term which comes from presence of gravitational waves.
Examining the various regimes for the free parameters $l_1, l_2 $ and $ v_{0x} $, we concluded that in $ f_{\times} $ polarization of the GW by changing the parameters in two cases of $ i) $ $l_1\gg 1, l_2\ll 1$ and $v_{0x}\gg 1$ and ii) $l_1\ll 1, l_2\gg 1$ and $v_{0x}\ll 1$ the GW effect may become large enough to hope to be visible.
\section{Acknowledgment}
Mohammad A. Ganjali would like to thanks the Kharazmi university
for supporting the paper with grant.

\end{document}